\documentclass[doublecol,A4paper]{epl2}

\usepackage{graphicx}
\usepackage{amssymb}
\usepackage{booktabs}
\usepackage{multirow}
\usepackage{caption}
\usepackage{mydefs}

\title{From Cooper-pair glass to unconventional
superconductivity\,:\\a unified approach to cuprates and pnictides}
\shorttitle{From Cooper-pair glass to unconventional SC}

\author{William Sacks\inst{1} \and Alain Mauger\inst{1} \and Yves Noat\inst{2}}
\shortauthor{W. Sacks, A. Mauger \& Y. Noat}

\institute{\inst{1} Institut de Min\'{e}ralogie, de Physique des
Mat\'{e}riaux, et de Cosmochimie (IMPMC), UMR 7590,

\inst{2} Institut des Nanosciences de Paris (INSP), UMR 7588,\\

Sorbonne Universit\'{e}s, UPMC Paris 6, \\
4 place Jussieu, 75252 Paris Cedex 05, France }

\pacs{74.72.h}{First pacs description} \pacs{74.20.Mn}{Second pacs
description} \pacs{74.20.Fg}{Third pacs description}

\abstract{We report a microscopic model wherein the unconventional
superconductivity emerges from an incoherent `Cooper-pair glass'
state. Driven by the pair-pair interaction, a new type of quasi-Bose
phase transition is at work. The interaction leads to the
unconventional coupling of the quasiparticles to excited pair
states, or `super-quasiparticles', with a non-retarded
energy-dependent gap. The model describes quantitatively the
quasiparticle excitation spectra of both cuprates and pnictides,
including the universal `peak-dip-hump' signatures, and for the
pseudogap phase above $T_c$. The results show that instantaneous
pair-pair interactions account for the SC condensation without a
collective mode.}

\begin{document}

\maketitle

Despite its wide applications, the BCS theory ~\cite{PR_BCS1957}
fails to account for the physical properties of a large variety of
high-$T_c$ superconductors (SC), the cuprate family, but also the
more recent iron-based superconductors. A striking feature of these
materials is the proximity to an insulating phase, whether
anti-ferromagnetic (cuprates), spin density wave (iron based SC,
Bechgard salts) or localization (ultra-thin films). Just beyond the
insulating phase, the SC dome appears in the phase diagram as a
function of carrier concentration between two critical points.
Understanding the transition from such an insulating to SC state is
still a major challenge.

Microscopic measurements reveal an unconventional quasiparticle (QP)
dispersion, the `peak-dip-hump' structure
\cite{RevModPhys_Fisher2007}, often attributed to the coupling to a
collective
mode\,\cite{AdvPhys_Eschrig2006,PRB_Berthod2013,PRL_Ahmadi2011,Chilett_Jing2015,PRL_Chi2012,PRL_Fasano2010}.
Although the peak to dip energy follows both the neutron resonance
and $T_c$ as a function of doping \cite{Trends_Zaza2002,
CurrOp_Song2013,Chilett_Jing2015}, the finer shape of the QP spectra
and their temperature dependence remain a challenge. Moreover, in
the temperature range [$T_c$,\,$T^*$] a pseudogap (PG) state
persists, having a Fermi-level gap $\Delta_p$ much larger than the
critical energy scale $k_B\,T_c$ in cuprates (see
\cite{Nat_Hashimoto2014} and ref. therein) and also in iron-based SC
\cite{Nat_Xu2011,NJPhys_Kwon2012,PRB_Shimojima2014}.

\begin{figure}[t]
\centering \vbox to 5.0 cm{
\includegraphics[width=8.6 cm]{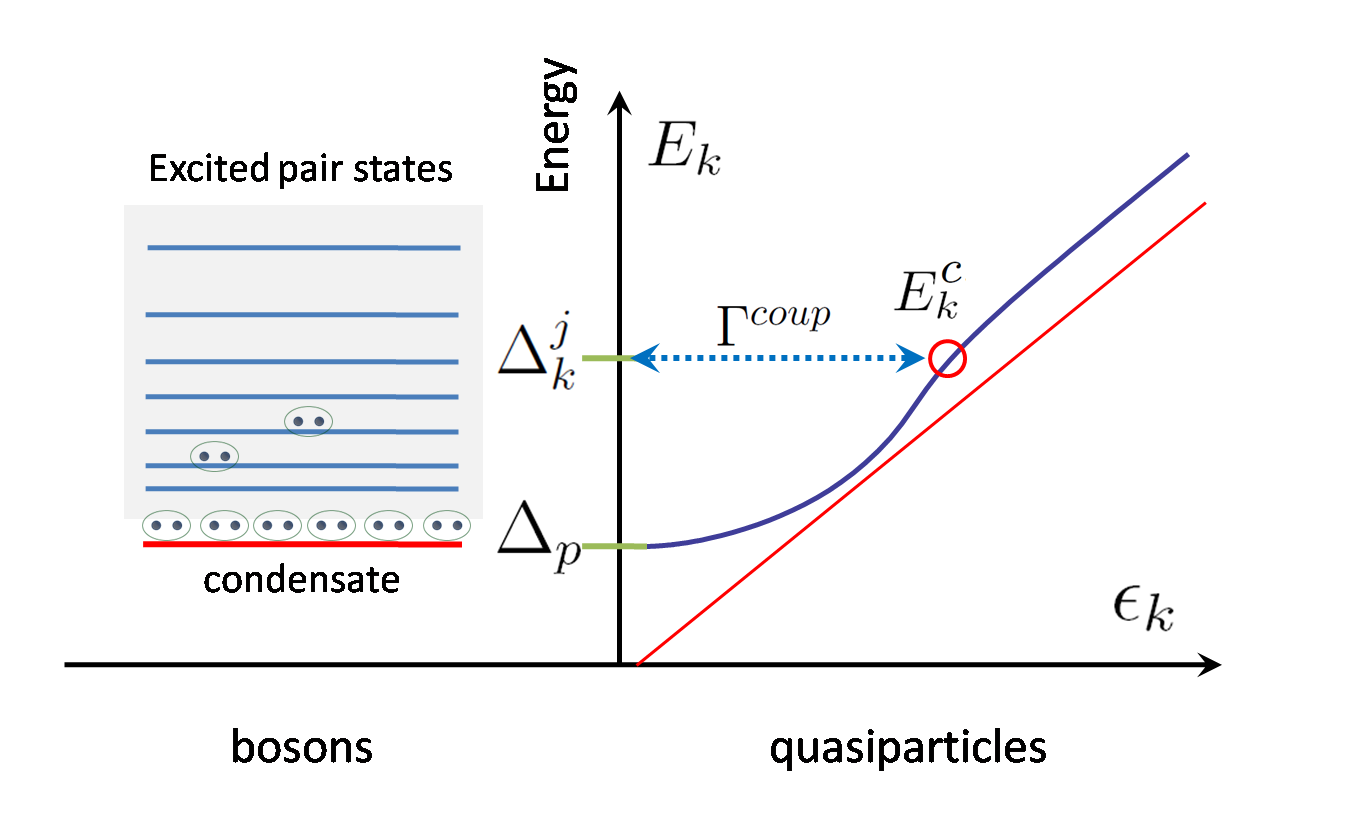}}
\caption{Illustration of the boson-fermion PPI model
\cite{SSciTech_Sacks2015}. The SC ground state (at $\Delta_p$) has
two distinct types of excitations\,: a distribution $\Delta_k^i$ of
pair (boson) excitations, left panel, and quasiparticle (fermion)
excitations. We demonstrate the strong coupling of the condensate QP
of energy $E^c_k$ to pre-existing excited pair states of equal
energy $\Delta_k^j$. The composite object is called a
`super-quasiparticle'.}\label{Fig1}
\end{figure}

In this letter, these questions are addressed within the pair-pair
interaction model (PPI). We show that the main unconventional
features of high-$T_c$ SC can be understood in a microscopic theory
wherein incoherent pairs in the {\it Cooper-glass state} interact to
form the coherent superconducting state. As a result of this novel
PPI, the quasiparticles become coupled to the excited pair states
(see Fig.\,\ref{Fig1}). These `super-quasiparticles' give rise to an
unconventional excitation spectrum wherein the gap function in the
SC state is energy dependent but non-retarded. The theory is in full
agreement with the experimental spectra on cuprates and pnictides,
despite the order of magnitude variation in the energy gap.

The results point to a universal mechanism in high-$T_c$ driven by
the interaction between pairs, giving key physical quantities such
as the condensation energy and elementary excitations, as a function
of temperature and doping. In particular, the `peak-dip-hump'
originates from instantaneous electron interactions, thus discarding
a bosonic mode as its origin in these materials.

\vskip 2mm

{\it Microscopic model}.
The hamiltonian describes normal electrons
coexisting with interacting preformed pairs\,:
\begin{equation}\label{ham1}
H = H_{0} + H_{pair}+H_{int}
\end{equation}
where the first term  $H_{0}$ describes the normal metal phase, and
the second term is the pairing hamiltonian:
\begin{equation}\label{hpair}
H_{pair} = -\sum_i\,\sum_k\ (\Delta^i_k\,\bkidag +
{\Delta^i_k}^*\,\bki)
\end{equation}
Here $\bkidag$ creates the $i$th pair state as composites of two
fermions\,: $\bki = \akidown\,\akiup$, and of binding energy
$\Delta^i_k$.

\begin{figure}[t]
\vbox to 5.8 cm{\centering
\includegraphics[width=8 cm]{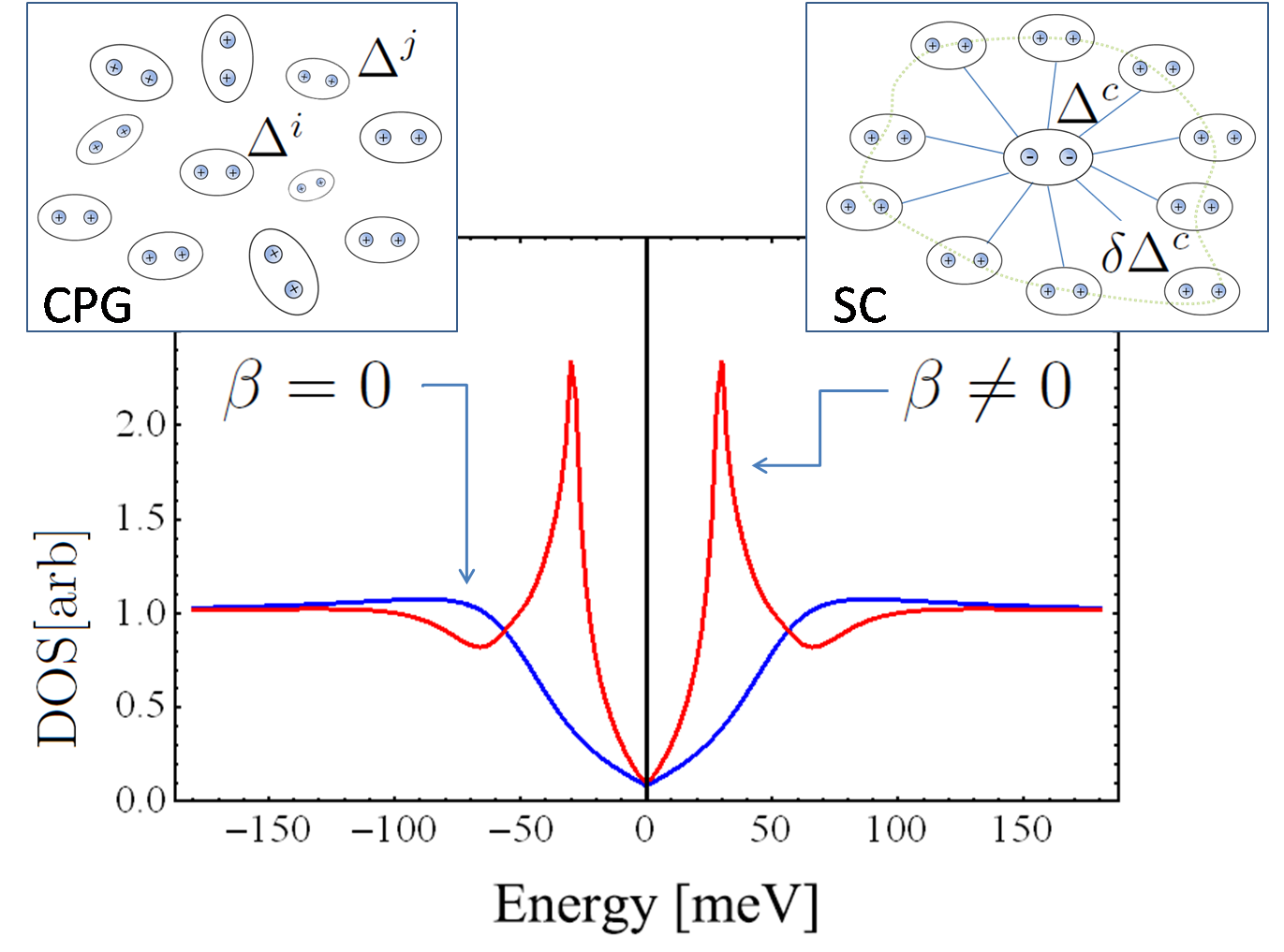}}
\caption{DOS in the Cooper-pair glass (CPG) state, blue curve,
showing a broad pseudogap of width $2\,<\Delta_i> = 2\,\Delta_0$ and
no coherence peaks. DOS in the SC state, red curve, with pronounced
dips due to the QP excited-pair coupling.}\label{Fig2}
\end{figure}

The first two terms $H_{PG} = H_{0} + H_{pair}$ describe a
non-superconducting state, a Cooper-pair glass having no global
phase, formed by the superposition of pairs in random states. SC
coherence is achieved due to the pair-pair interaction term\, giving
rise to the characteristic DOS (Fig.\,\ref{Fig2}, red curve):
\begin{equation}\label{Hint}
H_{int} = \frac{1}{2}\, \sum_{i\neq j} \sum_{k,k'}\, \beta^{i,j}_{k,k'}\
\bkpj\,\bkidag + h.c.
\end{equation}
where $\beta^{i,j}_{k,k'}$ are the coupling coefficients, which we
later tie to $\beta^c$, the SC order parameter.

\vskip 2mm

{\it Cooper-pair glass state}.
The accumulated results of photoemission \cite{PRL_Kanigel2008},
local tunneling experiments
\cite{PRL_Renner1998_Temp,PRL_Renner1998_B,Nat_Gomes2007} and normal
coherence length \cite{SciRep_Kirzhne2014} imply a scenario in
which, contrary to BCS theory, the pseudogap is linked to some form
of precursor pairing \cite{Nat_Emery1995}. The existence of
Fermi-surface arcs just above $T_c$, as seen using ARPES
\cite{PRL_Kanigel2008}, is further evidence. Without the PPI
($H_{int} = 0$) we consider that the system consists of incoherent
preformed pairs with an energy distribution\,:
\begin{equation}\label{P0}
P_0(\Delta^i) \propto \frac{\sigma^2}{(\Delta^i-\Delta_0)^2 +
\sigma_0^2}
\end{equation}
where $\Delta_0$ and $\sigma_0$ are the average gap and the
half-width, respectively.

In the spinor notation: $\tilde{a}^i_k = (\akiup,\akidowndag)$, the
equation of motion is\,: $ i\hbar\,\frac{d\,\tilde{a}^i_k}{dt} =
[\tilde{a}^i_k, H_{PG}] = H^i_{PG}\, \tilde{a}^i_k$, where
$H^i_{PG}$ is the effective matrix\,:
\begin{equation}\label{matrix}
H^i_{PG} = \left(
    \begin{array}{cc}
               \epsilon_k & -\Delta^i_k \\
               -\Delta^i_k & -\epsilon_k \\
             \end{array}
           \right)
\end{equation}
The latter is diagonal in the quasiparticle basis\,:
$\tilde{\gamma}^i_k = \Lambda^{i}_{k}\ \tilde{a}^i_k$ with
eigenvalues, $E^{i\pm}_k = \pm \sqrt{\epsilon_k^2 +
{\Delta^i_k}^2}$, leading to\,: $ H_{PG} = \sum_{i}\,\sum_{k}\
{\tilde{\gamma}^{i\,\,\dag}_k}\, (E^i_k\
\sigma_z)\,\tilde{\gamma}^i_k $, where $\sigma_z$ is the standard
Pauli matrix. In the continuum limit the spectral function
${A}_{PG}(k,E)$ acquires a significant width
\cite{PhysicaC_Sacks2014} and the $T = 0$ DOS becomes a
convolution\,:
\begin{equation}\label{dos2}
N_{PG}(E) = N_n(E_F)\,\int_0^\infty
\,d\Delta^i\,P_0(\Delta^i)\,\frac{E}{\sqrt{E^2-\Delta^{i\,\,2}}}
\end{equation}
As a result of the pair distribution, the coherence peaks in the DOS
are absent (blue curve, Fig.\,\ref{Fig2}), a key feature of the
incoherent {\it Cooper-pair glass}. This state is intimately related
to the pseudogap observed once SC coherence is lost, i.e. at $T_c$
or within a vortex core.

\begin{figure}[t]
\centering \vbox to 7 cm{
\includegraphics[width=8.8 cm]{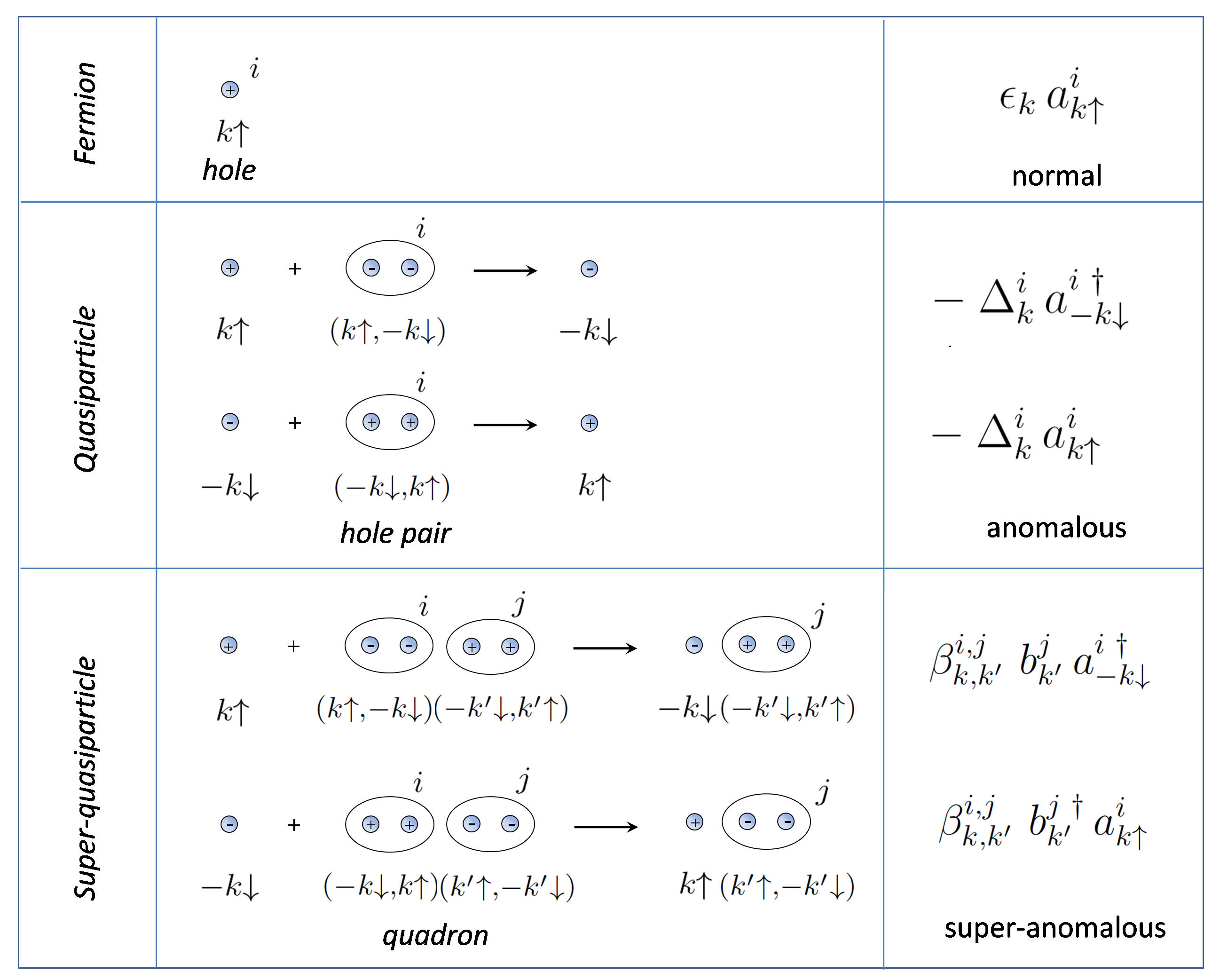}}
\caption{Summary of the various fermion/hole terms encountered in
the `exact' equation of motion (\ref{eq_motion}). Upper panel\,:
normal state electron/hole of energy $\epsilon_k$. Middle panel\,:
(anomalous term) standard BCS-type processes. Lower panel\,:
(super-anomalous term) an electron (hole) annihilates a {\it
quadron}, having zero spin and zero charge, leaving a {\it
super-quasiparticle} consisting of a pair plus a fermion. The
associated coupling energy is $\beta^{i,j}$.}\label{Fig3}
\end{figure}

\vskip 0.2cm {\it Equations of motion with $H_{int}$}. Adding the
term $[\tilde{a}^i_k, H_{int}]$ to the equation of motion, we
obtain\,:
\begin{eqnarray}\label{eq_motion}
i\hbar\, \dot a^i_{k\uparrow} &=& \epsilon_k\,\akiup -
\Delta^i_k\,\akidowndag + \sum_{j,k'}\null^{'}\, \beta^{i,j}_{k,k'}\
\bkpj\,\akidowndag \ \ \ \null \nonumber \\ i\hbar\, \dot
a^{i\,\,\dag}_{-k\downarrow} &=& - \epsilon_k\, \akidowndag -
\Delta^i_k\,\akiup + \sum_{j,k'}\null^{'}\, \beta^{i,j}_{k,k'}\
\bkpjdag\,\akiup \ \ \ \null
\end{eqnarray}
Obviously, without pairing ($\Delta^i = 0$), electrons are uncoupled
from holes, reflecting the normal state. To the contrary, the second
(anomalous) terms in (\ref{eq_motion}) are generated by the removal
of an electron-pair by a hole or a hole-pair by an electron
(Fig.\,\ref{Fig3}, middle panel). The third term is new\,: the final
state now contains a fermion triplet, which we call
`super-anomalous'. For a fixed ($j,k$), a {\it quadron} of zero spin
and charge is annihilated leaving a pair plus a fermion
(Fig.\,\ref{Fig3}, lower panel).

Since the $i$th electron (hole) is also coupled to all $j \neq i$,
the hamiltonian cannot be simply diagonalized in terms of a set of
quasiparticle operators \{$\tilde{\gamma}^i_k $\}. However, the
fermion operator triplet can be decoupled by the quantum average of
pair permutations\,:
\begin{eqnarray}\label{approx1 bis}
\bkpj\,\akidowndag &\simeq& <\akpjdown \akpjup>\,\akidowndag
     \\
 \null &+& <\akidowndag \akpjdown> \akpjup +
<\akpjup \akidowndag> \akpjdown \nonumber
\end{eqnarray}
resulting in the equation of motion\,:
\begin{eqnarray}\label{mfeq4}
i\hbar\, \frac{d\,\tilde{a}^i_k}{dt} &=& (H^i_{PG}\, + \,\delta\Delta^i_k\,\mathcal{J})\,\tilde{a}^i_k \\
&+& \sum_{j,k'}\null^{'}\,
\beta^{i,j}_{k,k'}\,\left[\Gamma^{i,j}_{k,k'}(\uparrow \uparrow)\,
\tilde{a}^j_{k'} + \Gamma^{i,j}_{k,k'}(\uparrow
\downarrow)\,\tilde{a}^{j\,\,\dag}_{k'} \right] \nonumber
\end{eqnarray}
in which the two $\Gamma$-matrix coefficients are\,:
$$\Gamma^{i,j}_{k,k'}(\uparrow \uparrow) =
\left(
  \begin{array}{cc}
    <\akidowndag \akpjdown> & 0 \\
    0 & <\akiup \akpjupdag> \\
  \end{array}
\right)
$$
$$\Gamma^{i,j}_{k,k'}(\uparrow \downarrow) =
\left(
  \begin{array}{cc}
    0 & <\akpjup \akidowndag> \\
    <\akpjdowndag \akiup> & 0 \\
  \end{array}
\right)
$$
and $\mathcal{J} = \left(
                   \begin{array}{cc}
                     0 & 1 \\
                     1 & 0 \\
                   \end{array}
                 \right) $.
These equations display coupling coefficients depending on different
states $(k,k')$ and different pairs $(i,j)$, where two are {\it spin
aligned}\,, $\Gamma(\uparrow \uparrow)$, and two are {\it spin
reversed}, $\Gamma(\uparrow \downarrow)$. The vertex amplitudes
imply instantaneous electron/hole interactions, all inherent to the
super-anomalous term of Fig.\,\ref{Fig3}. The new correction to the
gap function is\,:
\begin{equation}\label{gapcorrection}
    \delta\Delta^i_k = \sum_{j,k'}\null^{'}\, \beta^{i,j}_{k,k'} <\bkpj>
\end{equation}
which follows directly from (\ref{eq_motion}) with\,: $\bkj
\rightarrow \,<\bkj>$.

\vskip 0.2cm {\it Quasiparticle coupling}. In order to define
Bogoliubov quasiparticles in such a system, we apply the basis
transformation which diagonalizes the first two terms on the r.h.s.
of equation (\ref{mfeq4})\,: $$\overline \mathcal{O} = \overline
\Lambda^{i}_{k}\,\mathcal{O}\,\overline \Lambda^{i\,\,{-1}}_{k}$$
for a general operator $\mathcal{O}$. Writing the quasiparticle
basis as\, $\tilde \gamma^i_k = \overline \Lambda^{i}_{k}\
\tilde{a}^i_k$\,, yields\,:
\begin{equation}\label{mfterm2}
i\hbar\, \frac{d\,\tilde \gamma^i_k}{dt} = \overline
\Lambda^{i}_{k}\,(H^i_{PG} +
\,\delta\Delta^i_k\,\mathcal{J})\,\overline \Lambda^{i\,\,{-1}}_{k}
\,\tilde{\gamma}^{i}_k = \overline E^i_k\,\sigma_z\,\tilde \gamma^i_k
\end{equation}
It is important to stress that the eigenvalues ($\overline E^i_k$)
depend on  the {\it modified gap function}; their dispersion is\,: $
\overline E^i_k = \sqrt{\epsilon_k^2 + (\overline{\Delta}^i_k)^2}
$, where $\overline{\Delta}^i_k = \Delta^i_k - \delta\Delta^i_k$,
which we note is first order in $\beta$.

Using the $\overline \Lambda^{i}_{k}$ transformation, the equations
of motion (\ref{mfeq4}) can now be written in terms of QP
operators\,:
\begin{eqnarray}\label{mfeq5}
i\hbar\, \frac{d\,\tilde{\gamma}^i_k}{dt} &=& \overline E^i_k\,\sigma_z\,\tilde \gamma^i_k \\
&+& \sum_{j,k'}\null^{'}\, \beta^{i,j}_{k,k'}\,\left[\overline
\Gamma^{i,j}_{k,k'}(\uparrow \uparrow)\, \tilde{\gamma}^j_{k'} +
\overline \Gamma^{i,j}_{k,k'}(\uparrow
\downarrow)\,\tilde{\gamma}^{j\,\,\dag}_{k'} \right] \nonumber
\end{eqnarray}
As a result of the PPI, the second term of the full equation of
motion (\ref{mfeq5}) contains the coupling of the $i$th
quasiparticle to all other quasiparticles $j\neq i$, with the QP-QP
coupling proportional to the $\overline\Gamma$ coefficients. It
implies that quasiparticles interact via pair states and conversely
that pair states interact via quasiparticles\,: a novel QP-pair
vertex is thus revealed.

To illustrate the effect of the coupling we focus on the case where,
for wave vectors $k$ and $k'$, only two quasiparticles of energy
$E^i_k$ and $E^j_{k'}$ become degenerate (higher degeneracies are
possible) and the exact operators satisfy\,: $i\hbar\,
\frac{d\,\tilde \gamma^i_k}{dt} = i\hbar\, \frac{d\,\tilde
\gamma^j_{k'}}{dt} = E^{ex}_{kk'}\, \tilde \gamma^i_{k}$. In the
bi-spinor basis $ \tilde{\tilde{\gamma}}_k^j = ( \tilde{\gamma}_k^j,
\tilde{\gamma}_k^{j\,\,\dag})$, a new object of dimension 4 in the
$a_\mu$ fermions, the secular equation is obtained\,:
\begin{equation}\label{Secmatrix}
\left(
  \begin{array}{cc}
    (E^{ex}_{kk'}{\bf 1} - \overline E^i_k\,\sigma_z)\cdot {\bf 1}  &
    - \beta\ L^{i,j}_{k,k'} \\
    - \beta\ L^{j,i}_{k'k} &
    (E^{ex}_{kk'}{\bf 1} - \overline E^j_{k'}\,\sigma_z)\cdot {\bf 1} \\
  \end{array}
\right)
\times
\left(
\begin{array}{l}
\tilde{\tilde{\gamma}}_k^i \\
\tilde{\tilde{\gamma}}_k^{j\,\,\dag} \\
\end{array}
\right) = 0
                            \end{equation}
with $L^{i,j}$ a matrix of dimension 4\,:
\begin{equation}\label{Lmatrix}
L^{i,j}_{k,k'} = \left(
  \begin{array}{cc}
    \overline \Gamma^{i,j}_{k,k'}(\uparrow \uparrow) &
    \overline \Gamma^{i,j}_{k,k'}(\uparrow \downarrow) \\
    \overline \Gamma^{i,j}_{k,k'}(\uparrow \downarrow) &
    \overline \Gamma^{i,j}_{k,k'}(\uparrow \uparrow) \\
  \end{array}
\right)
\end{equation}
The analogy with the lowest order pairing matrix $H^i_{PG}$,
Eq.\,(\ref{matrix}), is striking. In the conventional BCS theory,
electron ($\epsilon_k$) and hole ($-\epsilon_k$) states are coupled
via the pair potential $\Delta$; here the PPI ($\sim \beta$) leads
to the coupling of the QP states ($\overline E^i_k$, $\pm \overline
E^j_{k'}$). Since the determinant of the secular matrix must vanish,
we obtain:
\begin{equation}\label{secequation}
(E^{ex\,\,2}_{kk'} - \overline E^{i}_k\null^2) (E^{ex\,\,2}_{kk'} -
\overline E^{j}_{k'}\null^2) = \beta^4\ {\rm \bf det} \left(
L^{j,i}\cdot L^{i,j} \right)
\end{equation}
where the explicit QP-QP coupling $\sim\,\beta^4 \Gamma^4$,
appearing on the r.h.s., is assumed to be small but finite.

The exact eigenstates $E^{ex}_{kk'}$ correspond to a new
super-quasiparticle, $(\tilde \gamma^i_k,\tilde \gamma^j_{k'})$, and
thus to the quadron
($\overline{\Delta}_k^i,\overline{\Delta}_{k'}^j$). While the
$\overline E^i_k$ are to first order in the PPI  $\propto \beta$,
the coupling involved in the super-quasiparticle is to higher order
in $\beta\,\Gamma$. Since the latter is small, the coupling of the
quasiparticles $\tilde \gamma^i_k$ and $\tilde \gamma^j_{k'}$ need
only be considered at the {\it degeneracy point}\,:
\begin{equation}\label{degen}
\overline E^{i}_k = \overline E^{j}_{k'} \end{equation} while
otherwise, $\overline E^i_k$ and $\overline E^j_{k'}$ are uncoupled.
This degeneracy condition thus plays a central role in the theory.

\vskip 0.2cm {\it Superconducting gap function}. The SC ground state
can be derived from the mean-field expression (\ref{gapcorrection})
wherein all pairs are assumed to be degenerate. The final-state gap
function is thus written\,:
\begin{equation}\label{gapbarSC}
    \overline{\Delta}_k = \Delta_{k,0} - \delta\Delta^{i=c}_k
\end{equation}
where $i=c$ indicates pairs of the condensate and $\Delta_{k,0}\,=$
$<\Delta^i_k>$. As in our previous work
Refs.\,\cite{SSciTech_Sacks2015}, we take the interaction to be
proportional to the DOS of preformed pairs\,: $ \beta^{i,j}_{k,k'} =
g_k \,g_{k'}\,P_0(\Delta_k^i)\,P_0(\Delta_{k'}^j) $, where $g_k$
takes into account the $d$-wave pairing. The crucial point is that
all the pairs $\Delta_{k'}^j$ are degenerate in the condensate.

\begin{figure}[t]
\centering \vbox to 10.6 cm{
\includegraphics[width=8.8 cm]{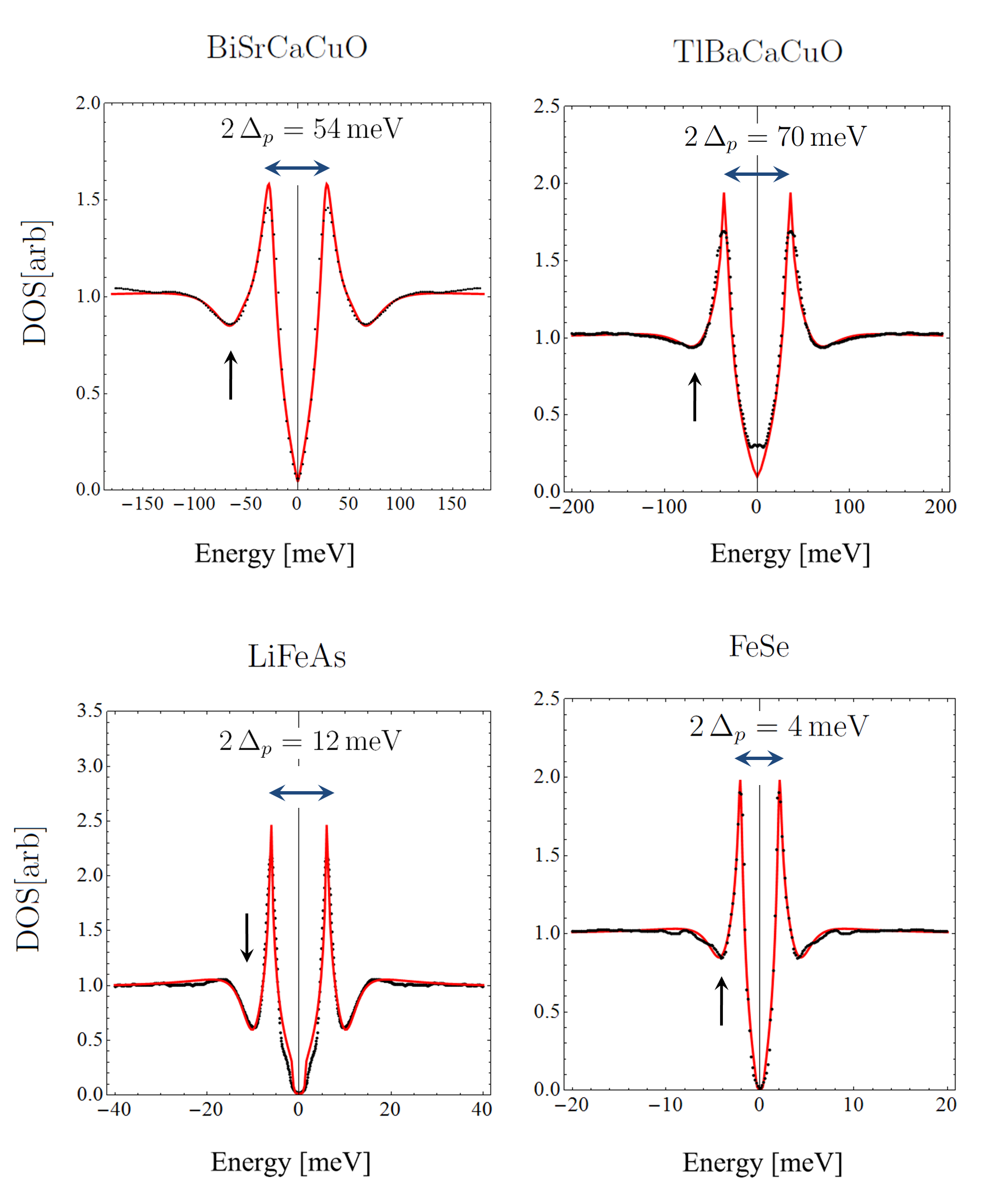}}
\caption{Fits to the tunneling DOS of 4 very different SC materials
using the same gap function (\ref{gapequation5}). We compare the
cuprate with iron-based materials\,: BiSrCaCuO (slightly overdoped
with $\Delta_p = 27$\,meV), TlBaCaCuO ($\Delta_p = 35$\,meV), LiFeAs
($\Delta_p = 6$\,meV), FeSe ($\Delta_p = 2$\,meV) taken from
Refs.(\cite{Sci_McElroy2005,JphysJap_Sekine2016,PRL_Chi2012,Sci_Song2011})
respectively. The other numerical values used for the fits
($\beta^c, \Delta_0, \sigma_0$) are summarized in Table\,I. The dip
position, indicated by the arrow in each case, follows
approximately\,: $E_{dip} \simeq \Delta_p + 2\,\beta^c$.
}\label{Fig5}
\end{figure}

The gap equation (\ref{gapbarSC}) must be self-consistent for zero
kinetic energy, wherein the QP states are at the Fermi level. Thus,
setting $\epsilon_k = 0$ and $\overline{\Delta}_k = \Delta_{k}^c$,
yields\,:
\begin{equation}\label{gapbarSC2}
    \Delta^{c}_k = \Delta_{k,0} - 2\,\beta_k^c\,P_0(\Delta^{c}_k)
\end{equation}
where $\beta_k^{c} = \frac{N_{oc}}{2}\,g_{k}\,
\sum_{k'}\,g_{k'}\,P_0(\Delta_{k'}^c)\,<b^c_{k'}>$ is the mean-field
{\it condensate} pair-pair interaction and $N_{oc}$, the number of
pairs ($N_{oc}\gg 1$). Since the mean-field parameter $\beta^c$ is
proportional to $N_{oc}(T)$, as a result of the quasi-Bose
transition, the second term represents the {\it condensation
energy}. As the temperature rises, it gradually decreases and
finally vanishes at $T_c$, contrary to the spectral
gap\,\cite{EPJB_Sacks2016} -- a clear departure from conventional
SC.

A key aspect of the problem is that the gap function $\Delta^{c}_k$
in equation (\ref{gapbarSC2}) must be modified for non-vanishing
kinetic energy, where a quasiparticle {\it becomes degenerate with
an excited pair state}\,(see Fig.\,\ref{Fig1}). The latter coupling
energy $\sim \beta^2\,\Gamma^2$ is to second order while the
renormalized gap function, proportional to $\beta^c$, remains large.
Thus, for excited states, $\epsilon_k
> 0$, the gap equation (\ref{gapbarSC}) is\,:
\begin{equation}\label{gapbarSC4}
    \overline{\Delta}^{i=ex}_k = \Delta_{k,0} - 2\,\beta_k^c\,P_0(\Delta^{i=ex}_k)
\end{equation}
where both $\Delta_{k,0}$ and $\beta_k$ are assumed independent of
$\epsilon_k$. Recalling equation (\ref{degen}), the {\it correct
degeneracy point} is $\Delta^{i=ex}_k = \overline E_k$ where we
identify $i \rightarrow \Delta^{i=ex}_k$ as the excited pair,
degenerate with the state $j \rightarrow \overline E_k=
\sqrt{\epsilon_k^2 + (\overline \Delta_k)^2}$ of the condensate.
Dropping the overbar, the full gap equation for excited states,
reads\,:
\begin{equation}\label{gapequation5}
     {\Delta}_k(E_k)=\Delta_{k,0} - 2\,\beta_k^c\,P_0(\sqrt{\epsilon_k^2
+ \Delta_k(E_k)^2})
\end{equation}
We thus have an energy dependent and self-consistent equation for
the gap function which leads to a strictly non-hyperbolic QP
dispersion and, most significantly, gives rise to the dip in the
spectral function (Fig.\,\ref{Fig2}). One can now identify its
physical origin\,: the strong coupling of the SC quasiparticle with
excited pair states.

\vskip 0.2cm {\it Comparaison with experiments}. The instantaneous
interactions in the hamiltonian imply that the DOS can be calculated
with {\it no retardation effects} in the Green's function. If no
quasiparticle lifetime effect is invoked, at $T = 0$, the DOS for
the $d$-wave condensate can be calculated by the standard formulae
using $\frac{\partial \epsilon_k}{\partial E_k}$\,:
\begin{eqnarray}\label{DOS}
&\null&{N}_{SC}^{d}(E) = \mathcal{N}_n(E_F)
\int_{0}^{2\,\pi}\frac{d\theta}{2\,\pi}\,\int_0^\infty d\epsilon_k\
\delta(E_k - E) \nonumber\\
& \null &=\mathcal{N}_n(E_F)
\int_{0}^{2\,\pi}\frac{d\theta}{2\,\pi}\, \left[\frac{E_k -
\Delta_k(E_k,\theta) \frac{\partial\Delta_k}{\partial
E_k}}{\sqrt{E_k^2 - \Delta_k(E_k)^2}}\right]_{E_k=E}
\end{eqnarray}
where $\mathcal{N}_n(E_F)$ is the normal DOS at the Fermi energy.

The SC DOS is thus proportional to the derivative of the gap
function (\ref{gapequation5}) wherein the peak-dip-hump is due to
the interaction term: $ \frac{\partial\,\Delta_k(E_k)}{\partial\,
E_k} = 2\,\beta^c_k\, \frac{d\,P_0(E_k)}{dE_k}$. The controlling
parameters are thus the pair-pair interaction $\beta^c$ and the
condensate pair number, $N_{oc}$, but the distribution $P_0(E_k)$
plays an essential role. Since the derivative has two extrema, the
first one reinforces the QP coherence peaks, giving them an
unconventional wide shape, while the second extremum produces the
dip \cite{EPL_Cren2000}. As in our previous
work\,\cite{SSciTech_Sacks2015}, the DOS (\ref{DOS}) can be used to
fit a wide variety of tunneling spectra of high T$_c$
superconductors with remarkably few parameters: $\beta^c$, the mean
pair-pair interaction, $\Delta_0$ and $\sigma_0$ which characterize
the distribution of pair states.


\begin{table}[t!]
\centering
 \resizebox{8.6 cm}{!}{
\begin{tabular}{@{}lccccc@{}}
 SC parameters &\null & BiSrCaCuO\ & TlBaCaCuO\ & LiFeAs\ & FeSe\ \\
  \hline \hline \\
spectral gap& $\Delta_p$ & 27        & 35        & 6 & 2 \\
pair-pair int. & $\beta^c$ & 19.5      & 11.5      & 2.4 & .95   \\
dist. maximum & $\Delta_0$ & 48.5      & 52.5      & 7.6 & 3.2   \\
dist. width& $\sigma_0$ & 24.5      & 32        & 4 & 2\\
\end{tabular}}
\caption{Numerical values (all in meV) obtained from the fits of
Fig.\,\ref{Fig5} for the 4 different materials indicated. Note the
order of magnitude difference between between BiSrCaCuO and FeSe and
yet the same basic parameters apply. In all cases, we find that
$\Delta_0 \sim \Delta_p + \beta^c$ and $E_{dip} \simeq \Delta_p +
2\,\beta^c$, where $E_{dip}$ is the dip position.} \label{my-label}
\end{table}


Among the cuprates, the tunneling characteristics of BiSrCaCuO
(2212) or BiSrCaCuO (2203) have been the most clearly established
(see \cite{RevModPhys_Fisher2007} and references therein). Much
success has recently been done on iron-based SC (see
\cite{CurrOp_Song2013} and references therein), such as BaKFeAs,
doped Fe(Se,Te) \cite{Chilett_Jing2015}, as well as LiFeAs
\cite{PRL_Chi2012} or FeSe \cite{Sci_Song2011} where typical spectra
are shown in Fig.\,\ref{Fig5}. Since we focus on the SC aspects of
the DOS, the background density is removed and the spectra
symmetrized, without affecting adversely the SC gap and
peak-dip-hump features. Along with a slightly overdoped BiSrCaCuO
\cite{Sci_McElroy2005}, Fig.\,\ref{Fig5} shows a recent high-quality
spectrum on a 3-layer TlBaCaCuO \cite{JphysJap_Sekine2016},
indicating the universality of the peak-dip-hump features. In the
same figure, the spectra are fitted using the PPI model.

The parameters of the fits are given in Table\,I. First, we note the
relatively sharp peaks at $e\,V = \pm \Delta_p$ in the iron-based SC
as compared to BiSrCaCuO and TlBaCaCuO. Indeed, higher peaks are
quite rare, partly due to thermal smearing at 4.2\,K but also due to
a finite quasiparticle lifetime, which we estimate to be $\sim
1.5$\, meV in the case of BiSrCaCuO and an order of magnitude less
for FeSe. The detailed shape of all the spectra are accurately
fitted using the same gap function\,(\ref{gapequation5}) in the DOS,
from energies within the gap, to the wide QP peaks and the
pronounced dip. The parameters thus have the same meaning despite
the range of values, and the very different composition and
structure of the materials.

We find that $E_{dip}-\Delta_p\simeq 2\beta^c$, where $\beta^c\simeq
2k_BT_c$ follows the SC dome but without the collective mode
scenario. Rather, it emerges from the novel QP-pair vertex inherent
to the super-anomalous term of the equation of motion
(\,\ref{eq_motion}). These super-quasiparticles cause the dip in the
spectrum and signal the long range SC order. The condensate PPI
energy $\beta^c$ depends on the product of the pairing amplitude
$\Delta_p$ with the carrier density $p$ : $\beta^c(p) \propto p
\times \Delta_p$, providing a simple explanation for the SC dome.
The mechanism is thus the interplay between the pair binding energy,
decreasing with $p$, and the number of pairs increasing with $p$.

\vskip 0.2cm {\it Conclusion}. We propose a scenario for high-$T_c$
superconductors wherein the initial incoherent state is the
Cooper-pair glass, whose properties explain the observed pseudogap
and Fermi-arc phenomena in agreement with both tunneling
\cite{PRL_Renner1998_Temp,JphysJap_Sekine2016,PhysicaC_Kawashima2010}
and ARPES \cite{Nat_Hashimoto2014} experiments. SC coherence results
from the novel pair-pair interaction, which adds a {\it quadron}
term to the hamiltonian giving rise to a new type of fundamental
excitation, the {\it super-quasiparticle}. The important effect is
the renormalized gap function, which is energy-dependent, but non
retarded.

The theory gives for the first time the correct temperature and
doping dependence of the quasiparticles in the SC to PG transition.
It reproduces quantitatively the experimental spectra of both
pnictides and cuprates, including the peak-dip-hump structure, and
attributes a common meaning to the fundamental parameters. In
conclusion, these features are not due to the coupling to a bosonic
mode, but rather emerge from instantaneous all-electron
interactions.

\newpage

\bibliography{library}

\end{document}